\documentclass[prl,twocolumn,superscriptaddress,letterpaper,nofootinbib,nopreprintnumbers,noshowpacs]{revtex4-1}

\usepackage{graphicx}
\usepackage{subfigure}
\usepackage{dcolumn}
\usepackage{bm}
\usepackage{hyperref}
\usepackage{epsfig}
\usepackage{color}
\usepackage[utf8]{inputenc}
\usepackage{comment}
\usepackage[all]{hypcap}
\usepackage{subfiles}
\usepackage{etoolbox}
\usepackage{amsmath}
\usepackage{amssymb}

\def\ba{\begin{eqnarray}}
\def\ea{\end{eqnarray}}
\def\bea{\begin{eqnarray}}
\def\eea{\end{eqnarray}}
\def\be{\begin{equation}}
\def\ee{\end{equation}}

\def\({\left(}
\def\){\right)}
\def\[{\left[}
\def\]{\right]}

\newcommand{\Msun}{\ensuremath{M_\odot}}
\newcommand{\MinSearchMass}{0.2 $M_\odot$}
\newcommand{\MaxSearchMass}{1.0 $M_\odot$}
\newcommand{\ObsOneStart}{September 12, 2015}
\newcommand{\ObsOneEnd}{January 19, 2016}
\newcommand{\SearchedMassRange}{2--600 $M_\odot$}
\newcommand{\STwoFourDates}{Feb 14, 2003 -- March 24, 2005}
\newcommand{\VolumeIncrease}{1000}
\newcommand{\ObsOneDataDuration}{48.16 days}
\newcommand{\BankMasses}{0.19 -- 2.0}
\newcommand{\BankMatch}{97\%}
\newcommand{\BankSize}{$500\thinspace332$}
\newcommand{\CoincTime}{20 ms}
\newcommand{\FRange}{45--1024 Hz}
\newcommand{\FLowSNR}{93.0\%}
\newcommand{\WaveformDuration}{470}

\newcommand{\LoudestFAR}{6.19 per year}
\newcommand{\NSims}{$374\thinspace480$}
\newcommand{\DMConstraintPointTwo}{33\%}
\newcommand{\DMConstraintOne}{5\%}
\newcommand{\RateError}{25\%}
\newcommand{\ZeroPointTwoRate}{\ensuremath {1.0 \times 10^{6} \, \mathrm{Gpc}^{-3} \mathrm{yr}^{-1}}}
\newcommand{\OneRate}{\ensuremath {1.9 \times 10^{4} \, \mathrm{Gpc}^{-3} \mathrm{yr}^{-1}}}
\newcommand{\compact}{ultracompact}
\newcommand{\DCCNumber}{LIGO-P1800158-v12}

\newtoggle{fullauthorlist}
\toggletrue{fullauthorlist}

\newtoggle{endauthorlist}
\toggletrue{endauthorlist}

\date{\today}
\begin{document}

\title{Search for sub-solar mass \compact{} binaries in Advanced LIGO's first observing run}

\iftoggle{endauthorlist}{
	%
	%
	\let\mymaketitle\maketitle
	\let\myauthor\author
	\let\myaffiliation\affiliation
	\author{The LIGO Scientific Collaboration}
     	\author{The Virgo Collaboration}
}{
	%
	%
	\iftoggle{fullauthorlist}{


\author{B.~P.~Abbott}
\affiliation{LIGO, California Institute of Technology, Pasadena, CA 91125, USA}
\author{R.~Abbott}
\affiliation{LIGO, California Institute of Technology, Pasadena, CA 91125, USA}
\author{T.~D.~Abbott}
\affiliation{Louisiana State University, Baton Rouge, LA 70803, USA}
\author{F.~Acernese}
\affiliation{Universit\`a di Salerno, Fisciano, I-84084 Salerno, Italy}
\affiliation{INFN, Sezione di Napoli, Complesso Universitario di Monte S.Angelo, I-80126 Napoli, Italy}
\author{K.~Ackley}
\affiliation{OzGrav, School of Physics \& Astronomy, Monash University, Clayton 3800, Victoria, Australia}
\author{C.~Adams}
\affiliation{LIGO Livingston Observatory, Livingston, LA 70754, USA}
\author{T.~Adams}
\affiliation{Laboratoire d'Annecy de Physique des Particules (LAPP), Univ. Grenoble Alpes, Universit\'e Savoie Mont Blanc, CNRS/IN2P3, F-74941 Annecy, France}
\author{P.~Addesso}
\affiliation{University of Sannio at Benevento, I-82100 Benevento, Italy and INFN, Sezione di Napoli, I-80100 Napoli, Italy}
\author{R.~X.~Adhikari}
\affiliation{LIGO, California Institute of Technology, Pasadena, CA 91125, USA}
\author{V.~B.~Adya}
\affiliation{Max Planck Institute for Gravitational Physics (Albert Einstein Institute), D-30167 Hannover, Germany}
\affiliation{Leibniz Universit\"at Hannover, D-30167 Hannover, Germany}
% ...

\collaboration{The LIGO Scientific Collaboration and the Virgo Collaboration}

\author{S.~Shandera}
\affiliation{The Pennsylvania State University, University Park, PA 16802, USA}


\maketitle


	}{
  		\author{The LIGO Scientific Collaboration}
  		\affiliation{LSC}
  		\email{lsc-spokesperson@ligo.org}
		\author{The Virgo Collaboration}
  		\affiliation{Virgo}
		\email{virgo-spokesperson@ego-gw.it}
	}
}


\begin{abstract}
\iftoggle{endauthorlist}{
}{
  \newpage
}

We present the first Advanced LIGO and Advanced Virgo search for \compact{}
binary systems with component masses between \MinSearchMass{} --
\MaxSearchMass{} using data taken between \ObsOneStart{} and \ObsOneEnd{}.  We
find no viable gravitational wave candidates.  Our null result constrains the
coalescence rate of monochromatic (delta function)
distributions of non-spinning (\MinSearchMass{}, \MinSearchMass{}) \compact{}
binaries to be less than \ZeroPointTwoRate{} and the coalescence rate of a
similar distribution of (\MaxSearchMass{}, \MaxSearchMass{}) \compact{}
binaries to be less than \OneRate{} (at $90\%$ confidence).  Neither black
holes nor neutron stars are expected to form below $\sim 1 M_\odot$ through
conventional stellar evolution, though it has been proposed that similarly low
mass black holes could be formed primordially through density fluctuations in
the early universe. Under a particular primordial black hole binary formation
scenario, we constrain monochromatic primordial black hole populations of 0.2
$M_\odot$ to be less than \DMConstraintPointTwo{} of the total dark matter
density and monochromatic populations of 1.0 $M_\odot$ to be less than
\DMConstraintOne{} of the dark matter density.  The latter strengthens the
presently placed bounds from micro-lensing surveys of MAssive Compact Halo
Objects (MACHOs) provided by the MACHO and EROS collaborations.

\end{abstract}

\pacs{Valid PACS appear here}


\maketitle

\section{\label{sec:intro}Introduction}

The era of gravitational wave astronomy began with the observation of the
binary black hole merger GW150914~\cite{Abbott:2016blz}.  Since then, four
additional binary black hole mergers~\cite{Abbott:2016nmj, Abbott:2017vtc,
Abbott:2017gyy, Abbott:2017oio} and one binary neutron star
merger~\cite{TheLIGOScientific:2017qsa} have been announced as of November 2017.
Thus far, Advanced LIGO and Advanced Virgo searches have targeted binary
systems with total masses from \SearchedMassRange{}~\cite{TheLIGOScientific:2016pea,
Abbott:2017iws}, but the LIGO and Virgo detectors are also
sensitive to \compact{} binaries with components below 1 $M_\odot$ if the
compactness (mass to radius ratio) is close to that of a black hole.  White
dwarf binaries, while often formed with components below one solar mass, are
not sufficiently compact to be a LIGO/Virgo gravitational wave source.  Neutron stars
or black holes are sufficiently compact as would be other exotic compact objects. Previous
gravitational wave searches for sub-solar mass \compact{} binaries used data from
initial LIGO observations from \STwoFourDates{}~\cite{Abbott:2005pf,
Abbott:2007xi}.  Advanced LIGO~\cite{TheLIGOScientific:2014jea} presently surveys a volume of space
approximately \VolumeIncrease{} times larger than the previous search for
sub-solar mass \compact{} objects therefore improving the chances of detecting
such a binary \VolumeIncrease{}-fold.

In conventional stellar evolution models, the lightest \compact{} objects are
formed when stellar remnants exceed $\sim 1.4 \Msun{}$, the Chandrasekhar mass
limit~\cite{Chandrasekhar:1935zz, Chandrasekhar:1931ih}.  Beyond the
Chandrasekhar mass limit, electron degeneracy pressure can no longer prevent
the gravitational collapse of a white dwarf. The lightest remnants that exceed
the Chandrasekhar mass limit form neutron stars ~\cite{glendenning2012compact}.
When even the neutron degeneracy pressure cannot prevent collapse, heavier
stellar remnants will collapse to black holes.  Some equations of state predict
that neutron stars remain stable down to $\sim 0.1 \Msun$~\cite{Potekhin:2013qqa}; there is no widely
accepted model for forming neutron stars below $\sim 1 M_\odot$, though
a recent measurement does not exclude the possibility of $0.92
M_\odot$ neutron star~\cite{Martinez:2017jbp}. Observationally, black holes
appear to have a minimum mass of $\sim 5\Msun$ with a gap between the observed
neutron star masses and black hole
masses~\cite{2010ApJ...725.1918O,2011ApJ...741..103F,2012ApJ...757...36K}.
Detecting \compact{} objects below one solar mass could challenge our ideas
about stellar evolution or possibly hint at new, unconventional formation
scenarios.

Beyond conventional stellar evolution, one of the most prolific black hole
formation models posits that primordial black holes (PBHs) could have formed in
the early universe through the collapse of highly over-dense
regions~\cite{Zeldovich1967,Hawking1971,Carr1974,Meszaros1974,Chapline1975}.
It has been suggested that PBHs could constitute a fraction of the missing dark
matter~\cite{Chapline1975}, though this scenario has been
constrained~\cite{Carr2016}. LIGO's detections have revived interest in black
hole formation mechanisms and, in particular, the formation of primordial black
holes (PBHs)~\cite{Bird:2016dcv, Sasaki:2016jop}.  Though there are proposals
on how to distinguish a primordial black hole distribution from an
astrophysical one~\citep{Kovetz:2016kpi}, disentangling them is challenging
when the populations overlap in mass. Hence, detection of sub-solar mass
\compact{} objects would provide the cleanest signature for determining
primordial formation. Still, recent proposals for non-baryonic dark matter
models can produce sub-solar mass black holes either by allowing a lower
Chandrasekhar mass in the dark sector~\cite{Shandera:2018xkn}, or by triggering
neutron stars to collapse into $\sim 1 M_\odot$ black holes
~\cite{Kouvaris:2018wnh}.

This letter describes a gravitational wave search for \compact{} binary systems
with component masses between \MinSearchMass{} and \MaxSearchMass{} using data
from Advanced LIGO's first observing run . No viable gravitational wave
candidates were identified. We briefly describe the data analyzed and the
anticipated sensitivity to sub-solar mass \compact{} objects, as well as the
search that was conducted, which led to the null result.  We then describe how
the null result constrains the merger rate of sub-solar mass binaries in the
nearby universe.  We consider the merger rate constraints in the context of
binary merger rate estimates most recently given by Sasaki et al~\cite{Sasaki:2016jop}
thereby constraining the fraction of dark matter density made up of PBHs
between \MinSearchMass{} and \MaxSearchMass{}.  Finally, we conclude with a
discussion of future work.

\section{Search}
\label{sec:search}
We report on data analyzed from Advanced LIGO's first observing run, taken from
\ObsOneStart{} -- \ObsOneEnd{} at the LIGO Hanford and LIGO Livingston
detectors. After taking into account data quality
cuts~\cite{TheLIGOScientific:2017lwt} and detector downtime, we analyzed a
total of \ObsOneDataDuration{} of Hanford-Livingston coincident data.
The data selection process was identical to that used in previous
searches~\cite{abbott2016binary}.

\begin{figure}
\includegraphics[width=\columnwidth]{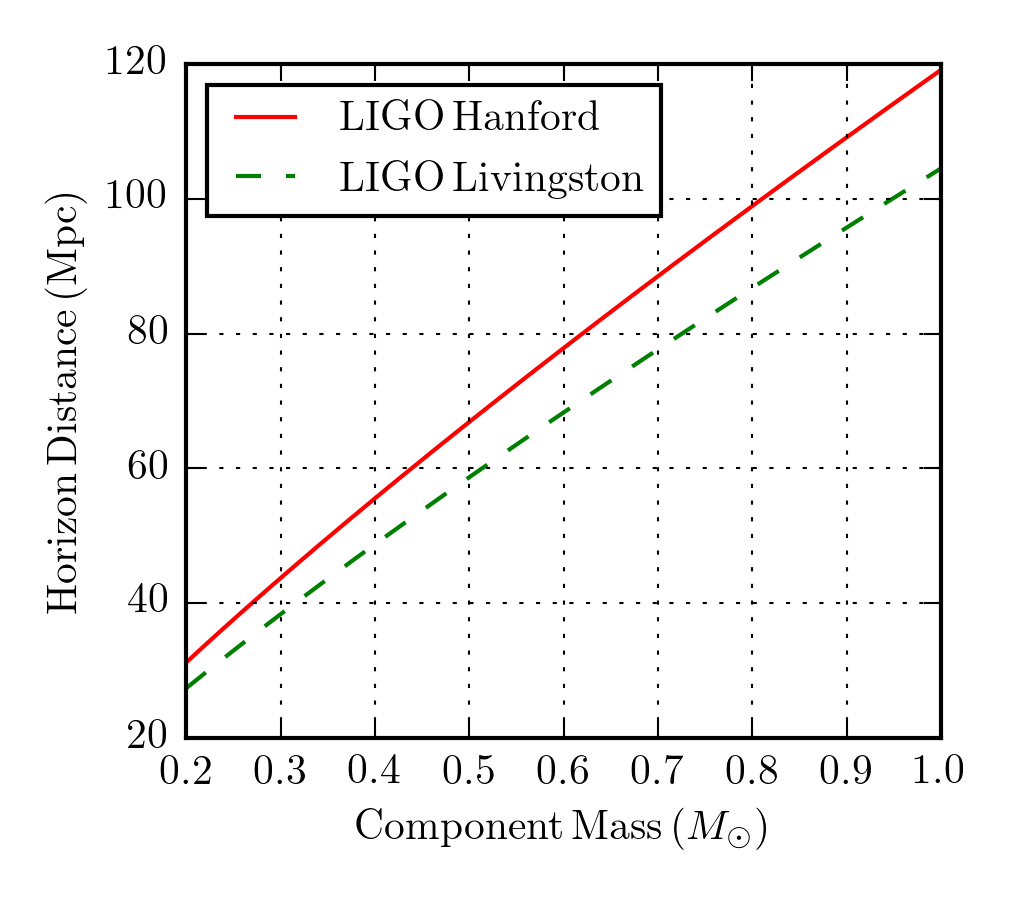}
\caption{\label{fig:horizon}
Distance to which an optimally oriented and aligned equal-mass \compact{} binary
merger would produce at least SNR 8 in each of the LIGO Livingston and LIGO
Hanford detectors as a function of component mass, based on the median
sensitivity obtained from our analyzed data.}
\end{figure}

During Advanced LIGO's first observing run, each LIGO instrument was sensitive
to sub-solar mass \compact{} binaries at extra-galactic distances. Figure
\ref{fig:horizon} shows the maximum distance to which an equal-mass compact
binary merger with given component masses would be visible at a signal-to-noise
ratio of 8 in either LIGO Hanford or LIGO Livingston.

The search was conducted using standard gravitational wave analysis
software~\cite{Cannon:2011vi,
Messick:2016aqy,gstlal,lalsuite,Ajith:2012mn,Capano:2016dsf}.  Our search
consisted of a matched-filter stage that filtered a discrete bank of templates
against the LIGO data.  The peak SNR for each template for each second was identified and
recorded as a trigger.  Subsequently, a chi-squared test was performed that
checked the consistency of the trigger with a signal~\cite{Messick:2016aqy}.
The triggers from each LIGO detector and gravitational wave template were
combined and searched for coincidences within \CoincTime{}.  Candidates that
pass coincidence were assigned a likelihood ratio, $\mathcal{L}$, that accounts
for the relative probability that the candidates are signal versus noise as a
function of SNR, chi-squared, and time delay and phase offset between
detectors.  Larger values of $\mathcal{L}$ were deemed to be more signal-like.
The rate at which noise produced candidates with a given value of $\mathcal{L}$
was computed via a Monte Carlo integral of the noise derived from
non-coincident triggers, which we define as the false alarm rate of candidate signals.

Our discrete bank of \BankSize{} template waveforms~\cite{SSMMethods2018}
conformed to the gravitational wave emission expected from general
relativity~\cite{PhysRevLett.74.3515, buonanno2009comparison}.  The bank
covered component masses in the detector frame between \BankMasses{} $M_\odot$
with \BankMatch{} fidelity.  While we restrict our analysis of the search
results to the sub-solar region, we have allowed for the possibility of high
mass ratio systems. Our template bank assumed that each binary component has
negligible spin.  Relaxing that assumption is a direction for future work, but
is a computationally challenging problem requiring resources well beyond those
used for this and previous LIGO analyses.  We integrated the template waveforms
between \FRange{}, with the longest waveform lasting about
$\WaveformDuration{}$ seconds.  Advanced LIGO is sensitive down to $\sim15$ Hz,
but integrating from that frequency would have been too computationally
burdensome.  Our choice to integrate from 45 Hz to 1024 Hz recovered \FLowSNR{}
of the total possible SNR that integration over the full band would have
provided.  Additional details are described in ~\cite{SSMMethods2018}.

No viable gravitational wave candidates were found.  Our loudest gravitational
wave candidate was consistent with noise and had a false alarm rate of
\LoudestFAR{}.

\section{Constraint on binary merger rate}
\label{sec:rates}

\begin{figure}
\includegraphics[width=\columnwidth]{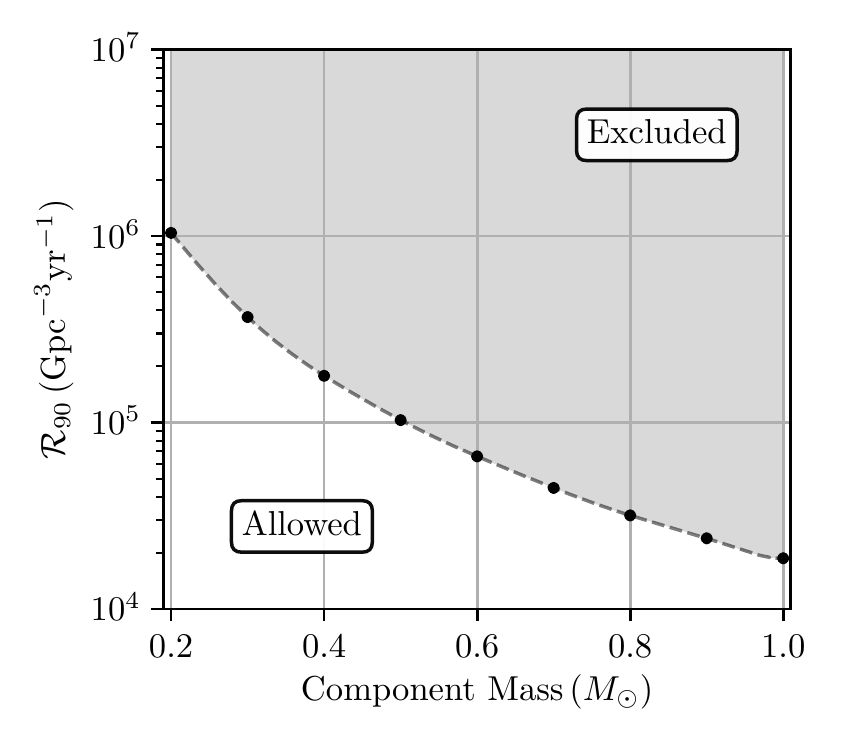}
\caption{\label{fig:rates}
Constraints on the merger rate of equal-mass \compact{} binaries
at the 9 masses considered. The gray region represents an exclusion
at 90\% confidence on the binary merger rate in units of $\text{Gpc}^{-3}\text{
yr}^{-1}$. These limits are found using the loudest event statistic formalism,
as described in section III and ~\cite{Biswas:2007ni}. The bounds presented
here are $\sim 3$ orders of magnitude stricter than
those found in initial LIGO’s search for sub-solar mass \compact{} objects
~\cite{Abbott:2005pf, Abbott:2007xi}
}
\end{figure}

We constrained the binary merger rate in this mass region by considering nine
monochromatic mass distributions with equal component masses and negligible spin.  We constructed
sets of simulated signals with component masses $m_i \in \{0.2, 0.3, \dots,
1.0\} M_\odot$ distributed uniformly in distance and uniformly on the sky.  We injected
 \NSims{} simulated signals into the LIGO data and conducted a
gravitational wave search with the same parameters as described in section
\ref{sec:search}.  We then calculated our detection efficiency as a function of
distance, $\epsilon_i(r)$. This allowed us to compute the volume-time, $\langle VT \rangle$,
that was accessible for our search via,
\begin{align}
\langle VT \rangle_i &= T \int 4\pi r^2 \epsilon_i(r) dr,
\end{align}
where $T$ is \ObsOneDataDuration{}. We then used the loudest event statistic
formalism~\cite{Biswas:2007ni} to compute an upper limit on the binary merger
rate in each mass bin to 90\% confidence,
\begin{align}
\mathcal{R}_{90,i} &= \frac{2.3}{\langle VT \rangle_i}. 
\end{align}
We report the upper limits on the binary merger rate in Fig. \ref{fig:rates}.
Several factors in our analysis could lead to uncertainty in
$\mathcal{R}_{90}$ at the \RateError{} level, including LIGO calibration errors
and Monte Carlo errors. However, these errors are far smaller than
potential systematic errors in the models we will be considering in the next
section, so we do not attempt to further quantify them in this work.

\section{Constraint on primordial black holes as dark matter}

The constraint on the binary merger rate places bounds on the total fraction of
dark matter made of primordial black holes, $f$. These bounds are derived from
the expected event rate for a uniform distribution of monochromatic PBHs with
mass $m_i$ as considered above. We follow a method originally proposed
by~\citep{Nakamura:1997sm,Ioka1998} and recently used to constrain $\sim 30
M_\odot$ PBH mergers by~\citep{Sasaki:2016jop}.

We assume an initial, early-universe, monochromatic distribution of PBHs. As the
universe expands, the energy density of a pair of black holes not too widely
separated becomes larger than the background energy density. The pair decouples
from the cosmic expansion and can be prevented from prompt merger by the local
tidal field, determined primarily by a third black hole nearest the pair. The
initial separation of the pair and the relative location of the primary
perturber determine the parameters of the initial binary. From those, the
coalescence time can be determined. Assuming a spatially uniform initial
distribution of black holes, the distribution of coalescence times for those
black holes that form binaries is
\begin{equation}
	dP = \begin{cases}
      	 \dfrac{3 \, f^{\tfrac{37}{8}}}{58} \left[f^{-\tfrac{29}{8}} \left(\dfrac{t}{t_c}\right)^{\tfrac{3}{37}} - \left(\dfrac{t}{t_c}\right)^{\tfrac{3}{8}} \right] \dfrac{dt}{t}, & \ t < t_c \\[15pt]
    	\dfrac{3 \, f^{\tfrac{37}{8}}}{58} \left[ f^{-\tfrac{29}{8}} \left(\dfrac{t}{t_c}\right)^{-\tfrac{1}{7}} -  \left(\dfrac{t}{t_c}\right)^{\tfrac{3}{8}} \right] \dfrac{dt}{t}, & \ t \geq t_c
    \end{cases}
\end{equation}
where $t_c$ is a function of the mass of the PBHs and the fraction of the dark matter they comprise:
 
\begin{equation}t_c = \frac{3}{170}\frac{c^5}{(Gm_i)^{5/3}} \frac{f^7}{(1+z_{\text{eq}})^4} \left(\frac{8 \pi}{3 H_{0}^2 \Omega_{\text{DM}}}\right)^{4/3}
\end{equation}

This expression is evaluated at the time today, $t_0$, then multiplied by
$n_{\text{BH}}$, the current average number density of PBHs, to get the model event
rate~\cite{Sasaki:2016jop}:
\begin{equation}
	\mathcal{R}_{model} = n_{\text{BH}} \left. \frac{dP}{dt} \right|_{t=t_0}.
\end{equation}

\begin{figure}
\includegraphics[width=\columnwidth]{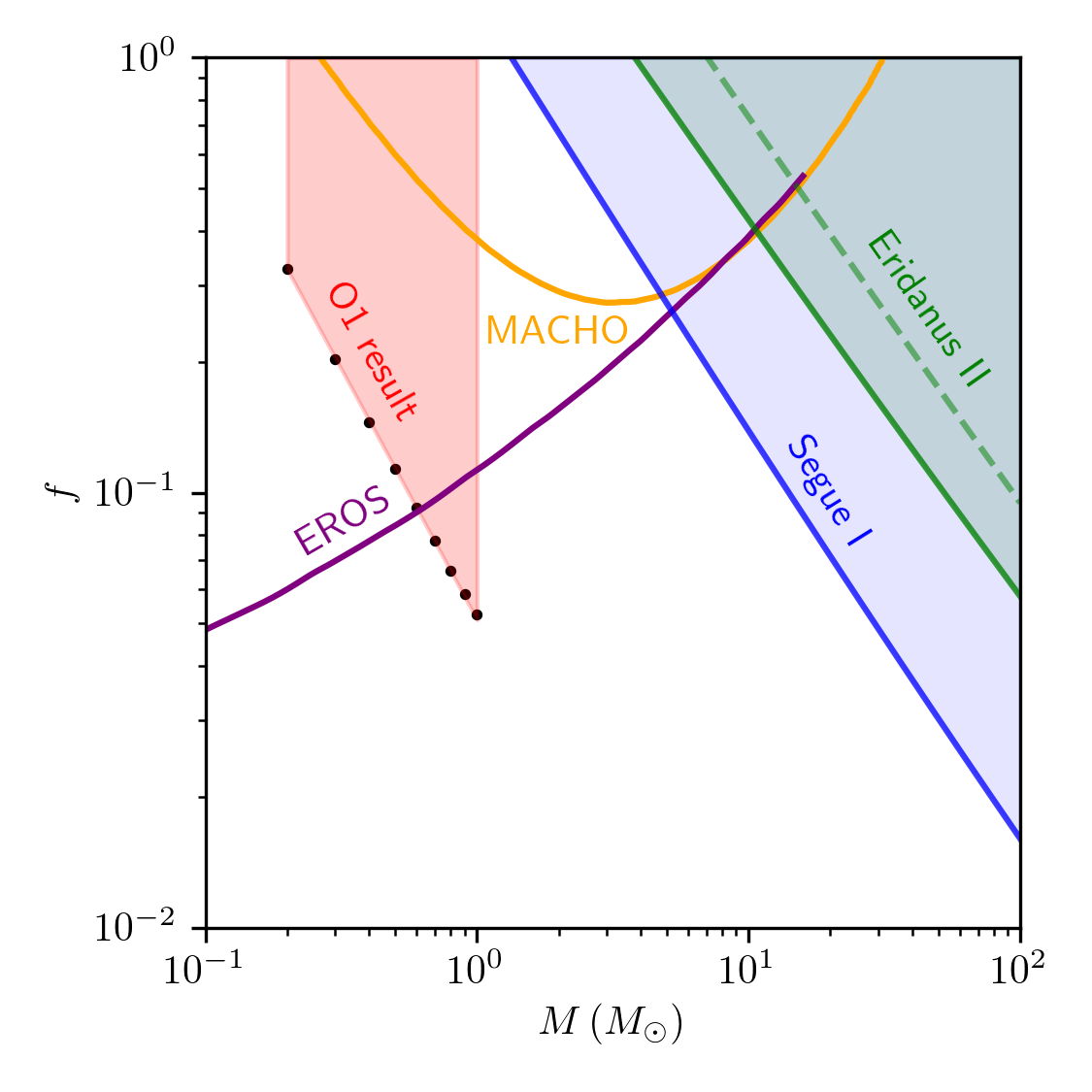}
\caption{\label{fig:fraction} Constraints on the fraction of dark matter
composed of primordial black holes for monochromatic distributions ($f =
\Omega_{\text{PBH}}/\Omega_{\text{DM}}$). Shown in black are the results for
the nine mass bins considered in this search. For this model of primordial black hole
formation, LIGO finds constraints tighter than those of the MACHO collaboration
~\cite{Allsman:2000kg} for all mass bins considered and tighter than the EROS
collaboration ~\cite{Tisserand:2006zx} for $m_i \in (0.7, 1.0) M_\odot$. The
curves shown in this figure are digitizations of the original results from
~\cite{Allsman:2000kg,Tisserand:2006zx,Koushiappas:2017chw,Brandt:2016aco}. We
use the Planck ``TT,TE,EE+lowP+lensing+ext'' cosmology~\cite{Ade:2015xua}.  }
\end{figure}

Given the measured event rate, $\mathcal{R}_{90,i}$, and a particular mass, the above expression can be inverted
to find a constraint on the fraction of dark matter in PBHs at that mass. The
results of this calculation using the measured upper limits on the merger rate
are shown in Fig. \ref{fig:fraction}. A discussion on how some assumptions of
this model may affect the constraints on $f$ shown in Fig.  \ref{fig:fraction},
are discussed in ~\cite{SSMMethods2018}.  The non-detection of a stochastic
background in the first observing run of Advanced
LIGO~\cite{TheLIGOScientific:2016dpb} also implies an upper limit on the merger
rate and therefore the PBH abundance. In particular, it is shown that the
non-detection of a stochastic background yields constraints that are about a
factor of two weaker than the targeted search~\cite{2016PhRvL.117t1102M,
Wang:2016ana, 2017JCAP...06..037C, 2017JCAP...09..037R}. 

\section{Conclusion}

We presented the first Advanced LIGO and Advanced Virgo search for \compact{} binary mergers with
components below 1 $M_\odot$.  No viable gravitational wave candidates were
found. Therefore, we were able to constrain the binary merger rate for
monochromatic mass functions spanning from 0.2 -- 1.0 $M_\odot$. Using a
well-studied model from the
literature~\cite{Nakamura:1997sm,Ioka1998,Sasaki:2016jop}, we constrained the
abundance of primordial black holes as a fraction of the total dark matter for
each of our nine monochromatic mass functions considered.

This work was only the first step in constraints by LIGO on new physics
involving sub-solar mass \compact{} objects.  The constraints
presented in Fig. \ref{fig:rates} (and consequently those that arise from the model
of binary formation we consider shown in Fig. \ref{fig:fraction}) may not apply if
the \compact{} binary components have non-negligible spin since the waveforms
used for signal recovery were generated only for non-spinning binaries. Future
work may either quantify the extent to which the present search could detect
spinning components, or expand the template bank to include systems with spin.
Third, we should consider more general distributions of primordial black hole
masses; extended mass functions allow for the possibility of unequal mass
binaries, and the effect of this imbalance on the predicted merger rate has not
been quantified.  We also stress that our present results do not rule out an
extended mass function that peaks below 0.2 $M_\odot$ and extends all the
way to LIGO's currently detected systems at or above 30 $M_\odot$.  Each model
would have to be explicitly checked by producing an expected binary merger rate
density that could be integrated against Advanced LIGO and Advanced Virgo search results.

The first two areas of future work are computational challenges. Lowering the
minimum mass and including spin effects in the waveform models could easily
increase the computational cost of searching for sub-solar mass \compact{}
objects by an order of magnitude each, which would be beyond the capabilities
of present LIGO data grid resources.  

Advanced LIGO and Advanced Virgo have not reached their final design sensitivities.  The distance to
which Advanced LIGO will be sensitive to the mergers of \compact{} binaries in
this mass range should increase by a factor of three over the
next several years~\cite{Aasi:2013wya}.  Furthermore, at least a factor of
ten more data will be available than what was analyzed in this
work.  These two facts combined imply that the merger rate constraint should
improve by $\gtrapprox 2$ orders of magnitude in the coming years.

\section{Acknowledgments}
The authors gratefully acknowledge the support of the United States
National Science Foundation (NSF) for the construction and operation of the
LIGO Laboratory and Advanced LIGO as well as the Science and Technology Facilities Council (STFC) of the
United Kingdom, the Max-Planck-Society (MPS), and the State of
Niedersachsen/Germany for support of the construction of Advanced LIGO 
and construction and operation of the GEO600 detector. 
Additional support for Advanced LIGO was provided by the Australian Research Council.
The authors gratefully acknowledge the Italian Istituto Nazionale di Fisica Nucleare (INFN),  
the French Centre National de la Recherche Scientifique (CNRS) and
the Foundation for Fundamental Research on Matter supported by the Netherlands Organisation for Scientific Research, 
for the construction and operation of the Virgo detector
and the creation and support  of the EGO consortium. 
The authors also gratefully acknowledge research support from these agencies as well as by 
the Council of Scientific and Industrial Research of India, 
the Department of Science and Technology, India,
the Science \& Engineering Research Board (SERB), India,
the Ministry of Human Resource Development, India,
the Spanish  Agencia Estatal de Investigaci\'on,
the Vicepresid\`encia i Conselleria d'Innovaci\'o, Recerca i Turisme and the Conselleria d'Educaci\'o i Universitat del Govern de les Illes Balears,
the Conselleria d'Educaci\'o, Investigaci\'o, Cultura i Esport de la Generalitat Valenciana,
the National Science Centre of Poland,
the Swiss National Science Foundation (SNSF),
the Russian Foundation for Basic Research, 
the Russian Science Foundation,
the European Commission,
the European Regional Development Funds (ERDF),
the Royal Society, 
the Scottish Funding Council, 
the Scottish Universities Physics Alliance, 
the Hungarian Scientific Research Fund (OTKA),
the Lyon Institute of Origins (LIO),
the Paris \^{I}le-de-France Region, 
the National Research, Development and Innovation Office Hungary (NKFI), 
the National Research Foundation of Korea,
Industry Canada and the Province of Ontario through the Ministry of Economic Development and Innovation, 
the Natural Science and Engineering Research Council Canada,
the Canadian Institute for Advanced Research,
the Brazilian Ministry of Science, Technology, Innovations, and Communications,
the International Center for Theoretical Physics South American Institute for Fundamental Research (ICTP-SAIFR), 
the Research Grants Council of Hong Kong,
the National Natural Science Foundation of China (NSFC),
the Leverhulme Trust, 
the Research Corporation, 
the Ministry of Science and Technology (MOST), Taiwan
and
the Kavli Foundation.
The authors gratefully acknowledge the support of the NSF, STFC, MPS, INFN, CNRS and the
State of Niedersachsen/Germany for provision of computational resources.
 Funding for this project was provided by the Charles E.  Kaufman Foundation of The Pittsburgh Foundation.  Computing resources and personnel for this project were provided by the Pennsylvania State University. This article has been assigned the document number \DCCNumber{}.

\bibliographystyle{unsrt}
\bibliography{O1ssm}
\newpage
\iftoggle{endauthorlist}{
  %
  %
  \let\author\myauthor
  \date[]{} 
  \let\affiliation\myaffiliation
  \let\maketitle\mymaketitle
  \title{Authors}
  \pacs{}
  
  \newpage
  \maketitle
}

\end{document}